# Non-equilibrium modeling of arc plasma torches

**J P Trelles, J V R Heberlein, and E Pfender**
Department of Mechanical Engineering, University of Minnesota, Minneapolis, MN 55455, USA

E-mail: jptrelles@me.umn.edu

**Abstract.** A two-temperature thermal non-equilibrium model is developed and applied to the three-dimensional and time-dependent simulation of the flow inside a DC arc plasma torch. A detailed comparison of the results of the non-equilibrium model with those of an equilibrium model is presented. The fluid and electromagnetic equations in both models are approximated numerically in a fully-coupled approach by a variational multi-scale finite element method. In contrast to the equilibrium model, the non-equilibrium model did not need a separate reattachment model to produce an arc reattachment process and to limit the magnitude of the total voltage drop and arc length. The non-equilibrium results show large non-equilibrium regions in the plasma – cold-flow interaction region and close to the anode surface. Marked differences in the arc dynamics, especially in the arc reattachment process, and in the magnitudes of the total voltage drop and outlet temperatures and velocities between the models are observed. The non-equilibrium results show improved agreement with experimental observations.

Keywords: thermal plasma, plasma torch, non-equilibrium, equilibrium, finite element method, multi-scale.

## 1. Introduction

Several thermal plasma applications are characterized by strong interactions between the plasma and a stream of cold-flow [1-2]. In these applications, large regions can exist where kinetic equilibrium between heavy-particles and electrons is not achieved, not only near the electrodes, but also in the plasma – cold-flow interaction regions. Thermal (or kinetic) non-equilibrium in those regions is due to the imbalance between transport and kinetic equilibration processes. To model the plasma in these applications more accurately, the Local Thermal Equilibrium (LTE) assumption needs to be abandoned in favor of a thermal non-equilibrium description in which the heavy-particles and electrons are assumed to have separate Maxwellian distributions [1].

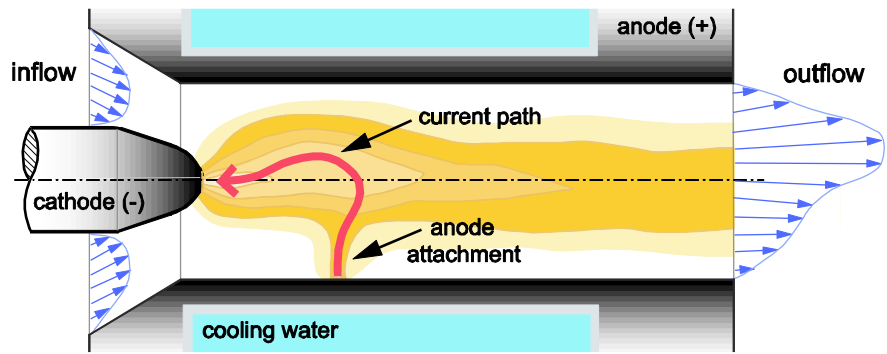

**Figure 1.** Schematic representation of the flow inside a DC arc plasma torch.

A typical application where strong plasma – cold-flow interactions are present is plasma spraying, one of the most versatile and widely used thermal plasma coating processes [1-3]. The occasional limited reproducibility found in plasma spraying processes is in great part due to our lack of understanding of the dynamics of the arc inside the spraying torch, the main component in these processes. The need to improve plasma spraying as well as other thermal plasma technologies has motivated the development of computational models capable of describing the arc dynamics inside plasma torches. This paper presents the development and application of a thermal non-equilibrium model applied to the simulation of the flow inside a direct current (DC) arc plasma torch typically used in plasma spraying. Figure 1 shows a schematic representation of the flow inside a DC, non-transferred arc plasma torch. This flow, despite the axi-symmetry and steadiness of the geometry and boundary conditions, is inherently three-dimensional and time-dependent, and thermodynamically can be considered as a dissipative structure in the sense of Prigogine [4]. The movement of the arc inside the torch forces the plasma jet, which promotes cold-flow entrainment and turbulence [2, 5].

The dynamics of the arc inside DC arc plasma torches are a result of the imbalance between flow drag forces and electromagnetic (or Lorentz) forces [6-7]. Because the total voltage drop across the torch is approximately linearly dependent on the arc length, the variation of the total voltage drop over time gives an indication of the arc dynamics inside the torch, which led to the identification of three distinct modes of operation of the torch: Steady, takeover, and restrike [6-9]. The flow inside a torch can change from the steady mode, to the takeover, and then to the restrike mode as the mass flow rate is increased or as the total current is decreased. The total voltage drop signal in these transitions between modes changes from steady, to periodic, and to quasi-periodic or chaotic, respectively [8-9].

A previous publication by the authors [10] describes *a reattachment model* which, when added to the LTE simulation of the flow inside an arc plasma torch, allows the mimicking of the breakdown process which leads to the formation of a new arc attachment. The results reported in [10] indicate some important limitations of the use of LTE models to describe the flow inside plasma torches (e.g. excessively high voltage drops), which point towards the importance of thermal non-equilibrium effects inside these torches.

The question of *how important non-equilibrium effects are* can only be answered properly by comparing the results of a non-equilibrium with those of an equilibrium model. Although equilibrium models are more convenient in terms of implementation and usage costs, they have important limitations, apart from their inability to describe non-equilibrium effects, which limit their usefulness for describing thermal plasma processes. One limitation of various LTE models is the need for a special treatment of the plasma-electrode interface to allow the passage of current. Another, as reported in [10], is the fact that results obtained with a LTE model are not capable of reproducing all the flow characteristics observed experimentally in DC plasma torches (i.e. observed voltage drop magnitudes, peak frequencies, and size of anode spot). It is, therefore, expected that the results of a non-LTE model would differ significantly from those of a LTE model.

Considering that the occurrence of non-equilibrium effects in thermal plasma processes is the rule rather than the exception [11], the need for non-equilibrium description of thermal plasmas is compelling. But, the added complexity of the equations of non-equilibrium models and of their thermodynamic and transport properties, and the subsequent significantly greater computational cost has limited the use of non-equilibrium models for the description of thermal plasma flows in realistic applications, which often involve three-dimensional and time-dependent characteristics. Although several non-equilibrium models have been applied to two-dimensional problems (e.g. [12-16]), and several equilibrium models have been applied to three-dimensional steady-state and time-dependent problems (e.g. [10, 17-25]), non-equilibrium models applied to three-dimensional problems are scarce. Furthermore, to the best knowledge of the authors, no non-equilibrium model applied to a three-dimensional and time-dependent problem has been reported in the literature yet.

Probably the most advanced non-equilibrium thermal plasma models to date are the works of Park [26] and Li *et al* [27-28], both applied to the three-dimensional steady-state modeling of an arc in cross flow. The arc in cross flow is a clear example with strong plasma – cold-flow interaction. Their models were developed for a three-component argon plasma (i.e. Ar, $Ar^+$, and $e^-$) in chemical non-equilibrium. Three-component plasma models were adequate in these studies due to the relatively low temperatures in the problem considered (i.e. < 15000 K). Due to the higher temperatures encountered inside plasma torches operating under typical operating conditions (i.e. ~ 25000 K), a three-component model of an argon plasma is not longer adequate and a four-component model (i.e. Ar, Ar$^+$, Ar$^{++}$ and e$^-$) needs to be used instead.

The non-equilibrium model presented here is based on a two-temperature four-component chemical equilibrium model of an argon plasma. Although chemical non-equilibrium effects could be important in the flow inside plasma torches, they have been neglected due to the increased computational cost associated with their description, especially due to the fact that four species need to be considered. To understand the importance of thermal non-equilibrium in plasma – cold-flow interactions, we compare the results from the non-equilibrium model with those from a thermal equilibrium model similar to the one developed in [10, 22-23]. The two-temperature chemical equilibrium model is referred here and thereafter as the NLTE (from Non-LTE) model, whereas the equilibrium model based on the LTE assumption is referred to as the LTE model. The NLTE model is equivalent to the LTE model if thermal equilibrium is assumed.

The results obtained with the NLTE model present significant differences from those obtained with the LTE model, especially in the observed arc dynamics, the total voltage drops, and outlet temperatures and velocities; the NLTE model results show improved agreement with experimental observations.

## 2. Mathematical models

### 2.1. Assumptions

The LTE and NLTE plasma models are based on the following assumptions:
- The plasma is considered as a continuum fluid.
- In the LTE model the plasma is characterized by a single temperature $T$, whereas in the NLTE model separate Maxwellian distributions are assumed for the heavy particles and for the electrons, characterized by a heavy-particle temperature $T_h$, and an electron temperature $T_e$, respectively.
- The plasma is in chemical equilibrium, and hence its composition is a function of pressure $p$ and $T$ only in the LTE model and of $p$, $T_h$, and $T_e$ in the NLTE model.
- The quasi-neutrality condition holds.
- Hall currents, gravitational effects, and viscous dissipation are considered negligible.

### 2.2. Thermal plasma equations

Along with the above assumptions, a thermal plasma flow is entirely described by 5 or 6 independent variables according to the LTE or NLTE model respectively. In the LTE model, we chose as unknowns the primitive variables: $p$, velocity $\vec{u}$, $T$, electric potential $\phi$, and magnetic vector potential $\vec{A}$ (a total of 9 components in a three-dimensional model). For the NLTE model, the variables used are: $p$, $\vec{u}$, $T_h$, $T_e$, effective electric potential $\phi_p$, and $\vec{A}$ (10 components in a three-dimensional model). These variables form the array $\mathbf{Y}$ of primitive variables, which for the LTE model is

$$\mathbf{Y} = \begin{bmatrix} p & \vec{u} & T & \phi & \vec{A} \end{bmatrix}, \tag{1}$$

whereas for the NLTE model is:

$$\mathbf{Y} = \begin{bmatrix} p & \vec{u} & T_h & T_e & \phi_p & \vec{A} \end{bmatrix}. \tag{2}$$

The superscript $t$ in (1) and (2) indicates "transpose".

The set of equations describing a thermal plasma flow can be expressed in compact form as a system of *transient-advective-diffusive-reactive* (TADR) equations in terms of **Y** as

$$\mathcal{R}(\mathbf{Y}) = \underbrace{\mathbf{A_0}\,\partial \mathbf{Y}/\partial t}_{transient} + \underbrace{(\mathbf{A}\cdot\nabla)\mathbf{Y}}_{advective} - \underbrace{\nabla\cdot(\mathbf{K}\nabla\mathbf{Y})}_{diffusive} - \underbrace{(\mathbf{S_1}\mathbf{Y} + \mathbf{S_0})}_{reactive} = \mathbf{0},$$

(3)

where $\mathcal{R}$ represents the vector of residuals and $\mathbf{A_0}$, $\mathbf{A}$, $\mathbf{K}$, $\mathbf{S_1}$, and $\mathbf{S_0}$ are matrices of appropriate sizes (i.e. for the NLTE model, $\mathbf{A_0}$ and $\mathbf{S_1}$ are of size 10x10, $\mathbf{A}$ 30x10, $\mathbf{K}$ 30x30, and $\mathbf{S_0}$ 10x1). These matrices are fully defined by the specification of the array of unknowns **Y** and a set of independent equations describing the plasma flow. Our LTE model is based on the equations of: (1) conservation of total mass; (2) conservation of mass-averaged momentum; (3) conservation of thermal energy; (4) conservation of electrical current; and (5) the magnetic induction equation. The NLTE model is based on the equations of: (1) conservation of total mass; (2) conservation of mass-averaged momentum; (3) conservation of heavy-particle thermal energy; (4) conservation of electron thermal energy; (5) conservation of electrical current; and (6) the magnetic induction equation. These equations are shown in Tables 1 and 2 as TADR systems of the form: *transient + advective − diffusive − reactive* = 0.

**Table 1.** Thermal plasma equations: LTE model.

| $i$ | $\mathbf{Y}_i$ | *transient* | *advective* | *diffusive* | *reactive* |
|---|---|---|---|---|---|
| 1 | $p$ | $\dfrac{\partial \rho}{\partial t}$ | $\vec{u}\cdot\nabla\rho + \rho\nabla\cdot\vec{u}$ | 0 | 0 |
| 2 | $\vec{u}$ | $\rho\dfrac{\partial \vec{u}}{\partial t}$ | $\rho\vec{u}\cdot\nabla\vec{u} - \nabla p$ | $-\nabla\cdot\vec{\tau}$ | $\vec{J}_q \times \vec{B}$ |
| 3 | $T$ | $\rho\dfrac{\partial h}{\partial t}$ | $\rho\vec{u}\cdot\nabla h$ | $-\nabla\cdot\vec{q}'$ | $\dfrac{Dp}{Dt} + \dot{Q}_J - \dot{Q}_r$ |
| 4 | $\phi$ | 0 | 0 | $-\nabla\cdot\vec{J}_q$ | 0 |
| 5 | $\vec{A}$ | $\dfrac{\partial \vec{A}}{\partial t}$ | $\nabla\phi - \vec{u}\times\nabla\times\vec{A}$ | $\eta\nabla^2\vec{A}$ | $\vec{0}$ |

**Table 2.** Thermal plasma equations: NLTE model.

| $i$ | $\mathbf{Y}_i$ | *transient* | *advective* | *diffusive* | *reactive* |
|---|---|---|---|---|---|
| 1 | $p$ | $\dfrac{\partial \rho}{\partial t}$ | $\vec{u}\cdot\nabla\rho + \rho\nabla\cdot\vec{u}$ | 0 | 0 |
| 2 | $\vec{u}$ | $\rho\dfrac{\partial \vec{u}}{\partial t}$ | $\rho\vec{u}\cdot\nabla\vec{u} - \nabla p$ | $-\nabla\cdot\vec{\tau}$ | $\vec{J}_q \times \vec{B}$ |
| 3 | $T_h$ | $\rho\dfrac{\partial h_h}{\partial t}$ | $\rho\vec{u}\cdot\nabla h_h$ | $-\nabla\cdot\vec{q}'_h$ | $\dfrac{Dp_h}{Dt} + \dot{Q}_{eh}$ |
| 4 | $T_e$ | $\rho\dfrac{\partial h_e}{\partial t}$ | $\rho\vec{u}\cdot\nabla h_e$ | $-\nabla\cdot\vec{q}'_e$ | $\dfrac{Dp_e}{Dt} - \dot{Q}_{eh} + \dot{Q}_J - \dot{Q}_r$ |
| 5 | $\phi_p$ | 0 | 0 | $-\nabla\cdot\vec{J}_q$ | 0 |
| 6 | $\vec{A}$ | $\dfrac{\partial \vec{A}}{\partial t}$ | $\nabla\phi_p - \vec{u}\times\nabla\times\vec{A}$ | $\eta\nabla^2\vec{A}$ | $\vec{0}$ |

In Tables 1 and 2, $\rho$ represents mass density, $t$ time, $\vec{\tau}$ the stress tensor, $\vec{J}_q$ current density, $\vec{B}$ magnetic field, $\vec{J}_q \times \vec{B}$ the Lorentz force; $h$, $h_h$, and $h_e$ are the equilibrium, heavy-particle, and electron enthalpies respectively (from here and thereafter, no subscript indicates an equilibrium property, while the subscripts $h$ and $e$ stand for heavy-particle and electron properties, respectively), $\vec{q}\,'$ total heat flux, $Dp/Dt$ is the pressure work with $D/Dt$ as the substantive derivative, $\dot{Q}_J$ the effective Joule heating, $\dot{Q}_r$ volumetric net radiation losses, $\dot{Q}_{eh}$ the electron − heavy-particle energy exchange term, $\eta = (\mu_0 \sigma)^{-1}$ the magnetic diffusivity with $\mu_0$ as the permeability of free space and $\sigma$ the electrical conductivity.

Consistency requires that a thermal non-equilibrium model must satisfy total thermal energy conservation; furthermore, if $T_h = T_e$ in this model, it should be reduced to a thermal equilibrium model. Our NLTE model is consistent with the LTE model because, considering that $h = h_h + h_e$ and $p = p_h + p_e$, adding lines 3 and 4 of Table 2 gives line 3 of Table 1, i.e. our thermal non-equilibrium model basically consists of a partition of total thermal energy between heavy-particles and electrons.

### 2.3. Plasma composition

The plasma is assumed to be in chemical equilibrium; therefore, according to the assumptions in Section 2.1, its composition is determined by the mass action law (minimization of Gibbs free energy), the quasi-neutrality condition, and Dalton's law of partial pressures [1]. For an argon plasma and the temperatures considered in this study, the plasma is primarily composed of 4 species, namely: argon atoms, ions, double ions, and electrons (i.e. Ar, Ar$^+$, Ar$^{++}$, and e$^-$), which are referred here by the sub-indexes 0, 1, 2, and $e$ respectively. It is important to emphasize the need to consider second ionization, because, as will be observed in the results shown in Section 5, temperatures near 30 kK are obtained inside the torch under certain operating conditions. This is in contrast with other non-equilibrium studies of thermal plasmas which needed to consider only first ionization (i.e. [26-28]).

Minimization of Gibbs free energy leads to the derivation of non-equilibrium Saha equations [1]. Even though two alternative forms of non-equilibrium Saha equations have commonly been used in the literature (i.e. see [29] and [12]), following [29], we have used Saha equations of the form:

$$\frac{n_e n_i}{n_{i-1}} = \frac{Q_e Q_i}{Q_{i-1}} \left( \frac{2\pi m_e T_e}{h_P^2} \right)^{\frac{3}{2}} \exp\left( -\frac{\varepsilon_i}{k_B T_e} \right) \qquad (4)$$

where the sub-index $i = \{1, 2\}$, $Q_i$ is the partition function of species $i$, $m_e$ the electron mass, $h_P$ is Planck's constant, $k_B$ is Boltzmann's constant, and $\varepsilon_i$ is the ionization energy of the $i$-times ionized atom. In equation (4) we have neglected the lowering of the ionization energy in order to reduce the computational cost of the calculation of plasma properties.

Dalton's law of partial pressures is stated as:

$$p = \sum_s p_s = \sum_s k_B n_s T_s \qquad (5)$$

where the summations include all the species (here and thereafter $\Sigma_s$ implies summation for $s = \{0, 1, 2, e\}$), $p_s$ is the partial pressure of species $s$, and $n_s$ the number density of species $s$. In the equilibrium case $T_s = T_h = T_e = T$, whereas in the non-equilibrium the temperatures of all heavy species are equal. Clearly $p = p_h + p_e$, where $p_h = \sum_{s=0}^{2} p_s$. The quasi-neutrality condition reads:

$$\sum_s Z_s n_s = 0 \quad \text{or} \quad n_e = \sum_{s=1}^{2} Z_s n_s \qquad (6)$$

where $Z_s$ is the charge number of species $s$, i.e. $Z_0 = 0$, $Z_1 = 1$, $Z_2 = 2$, and $Z_e = -1$.

Equations (4) to (6) allow the calculation of the plasma composition for a given total pressure and temperature(s). Figure 2 shows the composition of a 4-component 2-temperature argon plasma at 1 atm as a function of electron temperature for different values of the non-equilibrium parameter $\theta = T_e/T_h$. The plasma composition in the LTE model is given by the curves with $\theta = 1$. It can be seen in figure 2 that the concentration of double-ions becomes significant for temperatures above $\sim 22$ kK.

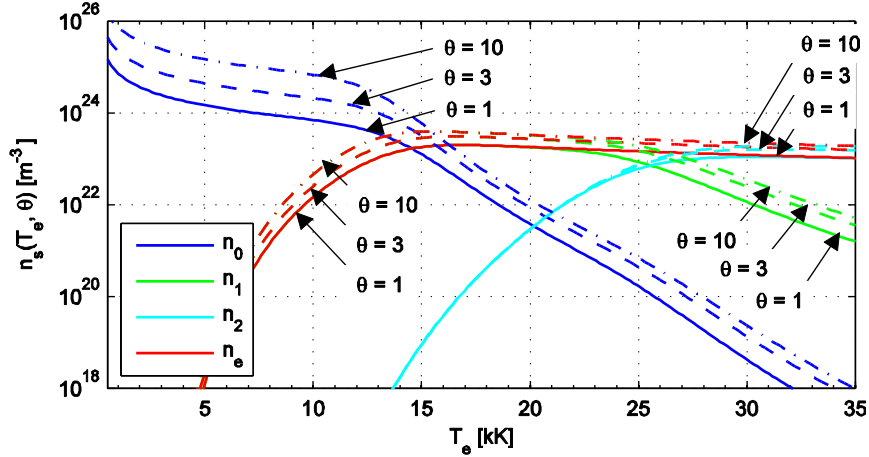

**Figure 2.** Chemical equilibrium composition of a 4-component 2-temperature argon plasma.

### 2.4. Thermodynamic and transport properties

Once the plasma composition is known, the thermodynamic properties total mass density, and the equilibrium, heavy-particle, and electron enthalpies are calculated by:

$$\rho = \sum_s m_s n_s \tag{7}$$

$$\rho h = \sum_s \tfrac{5}{2} k_B n_s T + \sum_{s=1}^{2} n_s \varepsilon_s \tag{8}$$

$$\rho h_h = \sum_{s=0}^{2} \tfrac{5}{2} k_B n_s T_h + \sum_{s=1}^{2} n_s \varepsilon_s \tag{9}$$

$$\rho h_e = \tfrac{5}{2} k_B n_e T_e \tag{10}$$

where $m_s$ is the mass and $\varepsilon_s$ the ionization potential of particle $s$, respectively. In expressions (8) and (9), the contributions to the enthalpy due to excitation of internal states as well as the lowering of the ionization energies have been neglected. Because of these definitions of enthalpies, chemical energy source terms are not needed in the thermal energy equations in Tables 1 and 2 (as used in [30]).

The accurate calculation of non-equilibrium transport properties of a plasma with more than 3 species requires significant computational resources. Therefore, to reduce the computational cost of the non-equilibrium calculations, a simplified, approximate, approach has been followed in this study, which consists on the use of look-up tables based on non-equilibrium transport properties at 1 atm reported in [12-13] and [31].

Figure 3 shows some of the thermodynamic and transport properties used in our models, where the properties for the LTE model correspond to the curves for $\theta = 1$. The marked dependence of the properties on $\theta$ for lower values of $\theta$ is evident. Enthalpies are significantly under-predicted and total mass density is over-predicted, at temperatures above $\sim 22$ kK, if these properties are calculated using only 3 species (i.e. Ar, Ar$^+$, e$^-$).

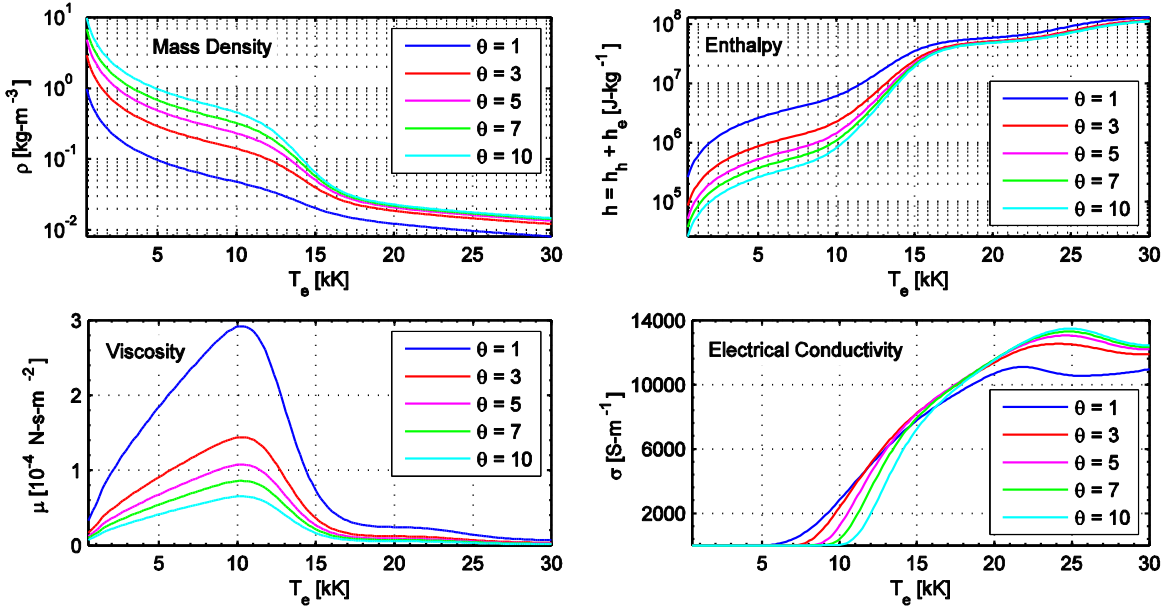

**Figure 3.** Thermodynamic and transport properties of a 2-temperature 4-component argon plasma.

The quasi-linear form of equation (3) requires expressing the terms involving $\rho$, $h$, $h_h$, and $h_e$ in Tables 1 and 2 in terms of the primitive variables $p$, $T$, $T_h$, and $T_e$. This is accomplished by determining the differentials of the former quantities with respect to the latter; specifically, differentiation of the total mass density and enthalpy gives for the LTE model:

$$\partial\rho = \rho_{,p}\,\partial p + \rho_{,T}\,\partial T \tag{11}$$

$$\partial h = h_{,p}\,\partial p + h_{,T}\,\partial T \tag{12}$$

whereas for the NLTE model:

$$\partial\rho = \rho_{,p}\,\partial p + \rho_{,T_h}\,\partial T_h + \rho_{,T_e}\,\partial T_e \tag{13}$$

$$\partial h_h = h_{h,p}\,\partial p + h_{h,T_h}\,\partial T_h + h_{h,T_e}\,\partial T_e \tag{14}$$

$$\partial h_e = h_{e,p}\,\partial p + h_{e,T_h}\,\partial T_h + h_{e,T_e}\,\partial T_e \tag{15}$$

where the sub-script "comma" denotes partial differentiation (i.e. $a_{,b} = \partial a/\partial b$). The term $h_{,T}$ in (12) is the specific heat at constant pressure $C_p$, whereas $h_{,p} = 1 + \partial\ln\rho/\partial\ln T|_p$ (see [32]). The coefficients accompanying the terms $\partial T_h$ and $\partial T_e$ in (13) and (14) can be interpreted as specific heats at constant pressure; as an example, the coefficient $h_{h,T_e}$ indicates the relative change of heavy-particle enthalpy due to an unitary change in electron temperature.

The volumetric heavy-particle specific heats $\rho h_{h,T_h}$ and $\rho h_{h,T_e}$, as well as a comparison between the calculated total volumetric specific heat at equilibrium conditions (summation of all the $\partial T_h$ and $\partial T_e$ coefficients in (14) and (15) for $\theta = 1$) with data from [1], all for $p = 1$ atm, are shown in figure 4. It can be observed in figure 4 the peaks due to the first and second ionization of argon, and in the left and middle frames, the relative importance of $\rho h_{h,T_e}$ with respect to $\rho h_{h,T_h}$ at temperatures above 10 kK. From these, it is clear that neglecting the Ar$^{++}$ species produces large errors for temperatures above $\sim$ 22 kK, and that the heavy-particle specific heat at constant pressure cannot be approximated by just the term $\partial h_h / \partial T_h |_p$. The good agreement between the total volumetric specific heat calculated in this study and the data obtained from [1] helps validate our thermodynamic properties calculations.

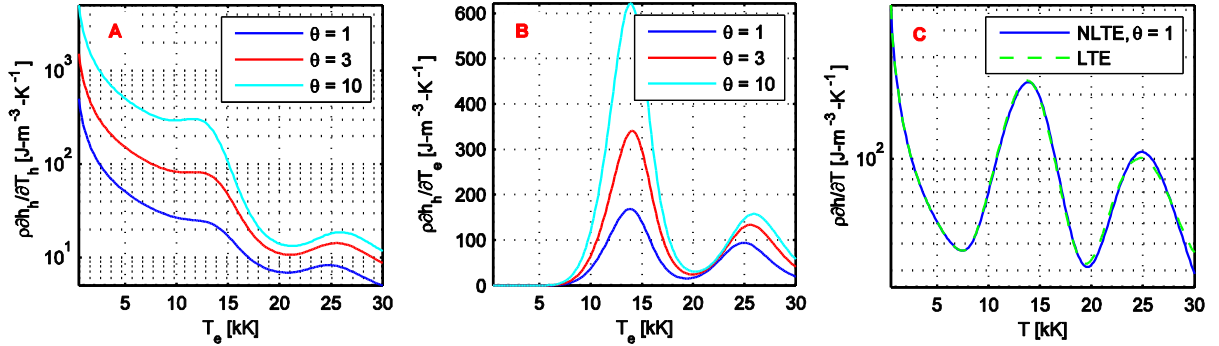

**Figure 4.** Volumetric specific heats at constant pressure. The LTE curve in frame **C** has been obtained with data from [1] and compared with the corresponding curve obtained in the calculations for $\theta = 1$.

### 2.5. Diffusive fluxes

The diffusive fluxes in Tables 1 and 2 are given by:

$$\bar{\bar{\tau}} = -\mu(\nabla \vec{u} + \nabla \vec{u}^t - \frac{2}{3}\nabla \cdot \vec{u}\,\bar{\bar{\delta}}) \qquad (16)$$

$$\vec{q}' = -\kappa \nabla T + h_e \vec{J}_e \qquad (17)$$

$$\vec{q}'_h = -\kappa_h \nabla T_h + \sum_{s \neq e} h_s \vec{J}_s \qquad (18)$$

$$\vec{q}'_e = -\kappa_e \nabla T_e + h_e \vec{J}_e \qquad (19)$$

where $\mu$ is the dynamic viscosity, $\bar{\bar{\delta}}$ the identity tensor, $\kappa$ thermal conductivity, which in the LTE model accounts for the translational and reactive thermal conductivities ($\kappa = \kappa_{translational} + \kappa_{reactive}$) (see [1, 33]), $\kappa_h$ and $\kappa_e$ are the translational heavy-particle and electron thermal conductivities respectively, $\vec{J}_s$ is the mass diffusion flux of species $s$ (and hence $\vec{J}_e$ is the mass diffusion flux of electrons). The current density $\vec{J}_q$ and $\vec{J}_e$ are related by:

$$\vec{J}_q \approx -\frac{e}{m_e}\vec{J}_e \qquad (20)$$

where $e$ represents the elementary charge unit. Equation (20) neglects the contribution of ion currents to the total current. Exploiting the fact that the chemical equilibrium assumption is used, equation (18) is simplified by defining a total heavy-particle thermal conductivity, which accounts for the translational and reactive components; therefore $\vec{q}'_h$ is approximated by:

$$\vec{q}'_h = -\kappa_{hr}\nabla T_h \qquad (21)$$

where $\kappa_{hr}$ is the translational-reactive heavy-particle thermal conductivity, which is obtained from [12].

*2.6. Electromagnetic relations*

The use of an effective electric potential $\phi_p$, as done in [14, 26-28], simplifies greatly a non-equilibrium model by permitting the formulation of the electromagnetic equations in terms of $\phi_p$ and $\vec{A}$ only. Specifically, if an effective potential is not used, the current density will depend on the electric field $\vec{E}$, on $\vec{A}$, and also on the different diffusion fluxes of the species in the plasma. In the LTE model, the electric potential is assumed equal to the effective one (i.e. $\phi \approx \phi_p$). Consistent with the definition of an effective electric potential, an effective electric field is defined according to:

$$\vec{E}_p = -\nabla\phi_p - \frac{\partial \vec{A}}{\partial t} \tag{22}$$

and is related to the real electric field $\vec{E}$ by:

$$\vec{E}_p \approx \vec{E} + \frac{\nabla p_e}{en_e}. \tag{23}$$

As shown in [26], expression (23) is an approximation of the effective electric field used in the Self-Consistent Effective Binary Diffusion (SCEBD) model developed for multi-component plasmas by Ramshaw and Chang [34]. The current density can then be expressed as:

$$\vec{J}_q = \sigma(\vec{E}_p + \vec{u} \times \vec{B}) \tag{24}$$

The magnetic vector potential is related to the magnetic field by:

$$\vec{B} = \nabla \times \vec{A} \tag{25}$$

In the derivation of the magnetic induction equation (i.e. equation 5 in Table 1 and equation 6 in Table 2), the gauge condition $\nabla \cdot \vec{A} = 0$ has been used.

*2.7. Source terms*

The effective Joule heating term is specified as:

$$\dot{Q}_J = \vec{J}_q \cdot (\vec{E} + \vec{u} \times \vec{B}) \tag{26}$$

whereas the real Joule heating term is given by $\vec{J}_q \cdot (\vec{E}_p + \vec{u} \times \vec{B})$; but, as clearly derived in [26], the effective heating mechanism is given by $\dot{Q}_J$ above. The difference between $\dot{Q}_J$ and the real Joule heating is equal to $\vec{U}_e \cdot \nabla p_e$, where $\vec{U}_e = \vec{J}_e/\rho_e$ is the diffusion velocity of electrons. Whereas the term $\vec{J}_q \cdot \vec{E}_p$ is always positive, the term $\vec{J}_q \cdot \vec{E}$ can be negative near the anode region. In that region, $\vec{U}_e \cdot \nabla p_e$ is usually positive, producing a net heating of the plasma. A more detailed analysis of the importance of the plasma heating mechanisms near the anode region is found in [26].

The net volumetric radiative energy loss is modeled according to [35], which considers only free-free and line radiation, and is given by:

$$\dot{Q}_r = 53.759\,\bar{n}_0\,\bar{n}_e^{1.25} + 910\,\bar{n}_e^{2} + 2572\,\bar{n}_e^{1.57} \tag{27}$$

where $\bar{n}_0 = 10^{-24} n_0$ and $\bar{n}_e = 10^{-20} n_e$ are dimension-less number densities, $n_e$ and $n_0$ are given in m⁻³, and $\dot{Q}_r$ is given in W·m⁻³. In expression (27), the first term represents electron-neutral free-free radiation; the second, electron-ion free-free radiation; and the third, $4p$-$4s$ line radiation. In this simple model, no absorption is taken into account, including resonant absorption of UV radiation, which would lead to additional heating of the plasma near the anode.

The electron – heavy-particle energy exchange term, which explicitly couples the heavy-particle and electron energy equations, is approximated by:

$$\dot{Q}_{eh} = \sum_{s\neq e} 3 k_B \frac{m_e}{m_s} n_e \nu_{es} (T_e - T_h)\delta_{es} \tag{28}$$

where $\nu_{es}$ is the collision frequency between electrons and species $s$ and is given by $\nu_{es} = n_s \bar{v}_e \overline{Q}_{es}$, with $\bar{v}_e$ as the thermal speed of electrons, and $\overline{Q}_{es}$ as the average collision cross section; and $\delta_{es}$ the inelastic collision factor, which, because we are dealing with a monatomic gas, is assumed equal to one.

Following Haidar [16], artificially increasing $\nu_{es}$ (and hence the energy exchange term) in regions with low electron temperature forces the rapid equilibration of $T_e$ with $T_h$, reducing significantly the stiffness of the electron energy equation in those regions. This procedure does not affect the results perceptibly, because for $T_e < 5000$ K, the number of electrons is negligible (see figure 2). We have followed a similar procedure by setting $n_e \leftarrow \max(n_e, 10^{18})$ ($n_e$ and $10^{18}$ in m⁻³) in (28). Figure 5 shows the electron – heavy-particle energy exchange term as a function of $T_e$ for different values of $\theta$ and for $p = 1$ atm. The net volumetric radiation loss term for $\theta = 1$ is also plotted in figure 5 as a reference, indicating that it is significantly smaller than the collisional energy-exchange, especially for high electron temperatures. The effect of limiting the minimum value of $n_e$ in (28) is observed for electron temperatures below ~ 5000 K.

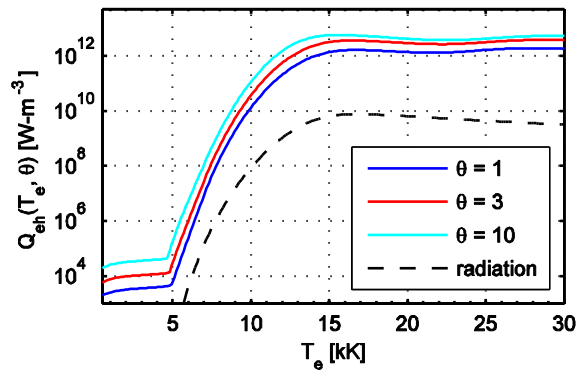

**Figure 5.** Electron – heavy-particle energy exchange term and volumetric radiation losses.

Considering that equation (28) can be written as $\dot{Q}_{eh} = K_{eh}(T_e - T_h)$, it can be noticed from equations 3 and 4 in Table 2 that, by letting $K_{eh} \rightarrow \infty$, the heavy-particle temperature is forced to be equal to the electron temperature, and hence the NLTE model will be forced towards the LTE model. It may appear that this fact could be employed to validate a non-equilibrium model with respect to an equilibrium one. But, as the boundary conditions for $T_e$ are usually different from those for $T_h$, this approach, generally, cannot be followed.

## 3. Numerical model

### 3.1. Multi-scale finite element method

To deal with the multi-scale nature of the thermal plasma equations given by (3), we use the Variational Multi-Scale Finite Element (VMS-FEM) method developed by Hughes and collaborators [36-37]. The VMS method uses a variational decomposition of a solution field $\mathbf{Y}$ into a large scale component $\overline{\mathbf{Y}}$ (which is described by the computational mesh) and a small or sub-grid scale component $\mathbf{Y'}$, whose effect into the large scales is modeled ($\mathbf{Y} = \overline{\mathbf{Y}} + \mathbf{Y'}$). The VMS finite element method applied to the TADR system given by equation (3) is given by:

$$\underbrace{\int_{\Omega} \overline{\mathbf{W}} \cdot \mathcal{R}(\overline{\mathbf{Y}})}_{\text{large scales}} - \underbrace{\int_{\Omega'} \mathcal{L}^* \overline{\mathbf{W}} \cdot \boldsymbol{\tau} \, \mathcal{R}(\overline{\mathbf{Y}})}_{\text{sub-grid scales}} = \mathbf{0} \tag{29}$$

where $\overline{\mathbf{W}}$ is the large scales weight function (typical finite element basis functions); $\Omega$ represents the spatial domain; $\Omega'$ represents the spatial domain minus the skeleton of the mesh formed by the elements boundaries; $\mathcal{L}^*$ the adjoint of the quasi-linear operator $\mathcal{L}$ defined from: $\mathcal{R}(\mathbf{Y}) = \mathcal{L}\mathbf{Y} - \mathbf{S_0}$ (see (3)); and $\boldsymbol{\tau}$ is the matrix of time scales, an algebraic approximation of an integral operator based on the element *Green's function* of the problem ($\boldsymbol{\tau} \approx \mathcal{L}^{-1}$) [36-37]. Equation (29) implies that $\mathbf{Y'} \approx -\boldsymbol{\tau}\,\mathcal{R}(\overline{\mathbf{Y}})$. To add robustness to the formulation, a *discontinuity-capturing-operator* is added to (29) [38]. More details of our implementation of the VMS-FEM applied to thermal plasmas are found in [22-23].

### 3.2. Solution approach

The spatial discretization of (29) leads to the formation of a very large system of time-dependent non-linear algebraic equations expressed as:

$$\mathbf{R}(t_h, \mathbf{X}_h, \mathbf{Y}_h, \dot{\mathbf{Y}}_h) = \mathbf{0} \tag{30}$$

where $\mathbf{R}$ represents the global array of residuals, the subscript $h$ indicates the discrete counterpart of a given continuous variable, $\mathbf{X}_h$ the computational domain, $\mathbf{Y}_h$ the global array of unknowns, and $\dot{\mathbf{Y}}_h$ its time derivative.

We discretize in time (30) using a fully-implicit predictor multi-corrector method [39], together with an automatic time-stepping procedure. The required solution of the non-linear system at each time step is accomplished by an inexact-Newton method together with a line-search globalization strategy. These techniques form an iterative procedure at each time step in which the values of an approximate solution and its residual vector at a given iteration $k$ are updated according to:

$$\left\| \mathbf{R}^k + \mathbf{J}^k \Delta \mathbf{Y}_h^{k+1} \right\| \le \gamma^k \left\| \mathbf{R}^k \right\| \quad \text{and} \quad \mathbf{Y}_h^{k+1} = \mathbf{Y}_h^k + \lambda^k \Delta \mathbf{Y}_h^{k+1} \tag{31}$$

where the super-index $k$ indicates the iteration counter, $\mathbf{J}$ is an approximation of the Jacobian matrix of the global system (30) ($\mathbf{J} \approx \partial \mathbf{R}/\partial \mathbf{Y}_h$), $\Delta \mathbf{Y}_h$ represents a correction to the actual solution, $\gamma$ is a parameter controlling the accuracy required for the linear solve in the current iteration, and $\lambda$ is the line-search parameter. The left-hand-side of (31) expresses the inexact-Newton condition whereas the right-hand-side the updating of the solution according to the line-search procedure. The approximate solution of a linear system required by the inexact Newton condition is accomplished by the preconditioned Generalized Minimal Residual (GMRES) method.

Although the NLTE model has only ~ 10% more unknowns, it typically is ~ 70% more expensive computationally than the LTE model. The additional cost comes from the added complexity of the TADR equations, the cost of the calculation of the non-equilibrium thermodynamic and transport properties, and on the overall increased non-linearity and stiffness of the NLTE model.

## 4. Computational domain and boundary conditions

The LTE and NLTE models have been used to simulate the plasma flow inside the commercial plasma torch SG-100 from Praxair Surface Technology, Concord, NH. Figure 6 shows the computational domain, which is formed by the region inside the torch limited by the cathode, the inflow region, the anode nozzle, and the torch exit (see figure 1). The center of coordinates is located in the center point of the cathode tip, and the diameter of the nozzle at the outlet is 8 mm. The computational domain is discretized using hexahedral finite elements in a mixed structured-unstructured mesh.

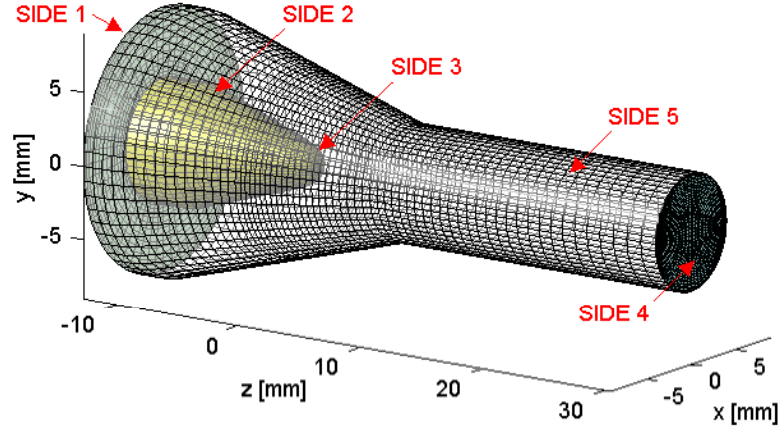

**Figure 6.** Computational domain, finite elements mesh, and boundary sides.

**Table 3.** Boundary conditions for the LTE and NLTE models [a, b].

| $Y_i$ | SIDE 1<br>inlet | SIDE 2<br>cathode | SIDE 3<br>cathode tip | SIDE 4<br>outlet | SIDE 5<br>anode |
|---|---|---|---|---|---|
| $p$ | $p_{,n} = 0$ | $p_{,n} = 0$ | $p_{,n} = 0$ | $p = p_{out}$ | $p_{,n} = 0$ |
| $\vec{u}$ | $\vec{u} = \vec{u}_{in}$ | $u_i = 0$ | $u_i = 0$ | $u_{i,n} = 0$ | $u_i = 0$ |
| $T$ | $T = T_{in}$ | $T = T_{cath}$ | $T = T_{cath}$ | $T_{,n} = 0$ | $-\kappa T_{,n} = h_w(T - T_w)$ |
| $T_h$ | $T_h = T_{in}$ | $T_h = T_{cath}$ | $T_h = T_{cath}$ | $T_{h,n} = 0$ | $-\kappa_{hr} T_{h,n} = h_w(T_h - T_w)$ |
| $T_e$ | $T_e = T_{in}$ | $T_e = T_{cath}$ | $T_e = T_{cath}$ | $T_{e,n} = 0$ | $T_{e,n} = 0$ |
| $\phi$ | $\phi_{,n} = 0$ | $\phi_{,n} = 0$ | $-\sigma\phi_{,n} = J_{cath}$ | $\phi_{,n} = 0$ | $\phi = 0$ |
| $\phi_p$ | $\phi_{p,n} = 0$ | $\phi_{p,n} = 0$ | $-\sigma\phi_{p,n} = J_{cath}$ | $\phi_{p,n} = 0$ | $\phi_p = 0$ |
| $\vec{A}$ | $A_i = 0$ | $A_{i,n} = 0$ | $A_{i,n} = 0$ | $A_{i,n} = 0$ | $A_{i,n} = 0$ |

[a] $a_{,n} = \partial a / \partial n$ differentiation in the direction of the (outer-) normal to the boundary.

[b] $i = x, y, z$ Cartesian coordinates defined in figure 6.

As seen in figure 6, the boundary of the computational domain is divided into 5 different *sides* to allow the specification of boundary conditions. Table 3 shows the boundary conditions used for the different variables in the LTE and NLTE models; these boundary conditions can be considered typical of simulations of arc plasma torches (i.e. see [10, 17-20, 22-23]).

In Table 3, $p_{out}$ represents the outlet pressure equal to 101.325 kPa (= $p_0$, atmospheric pressure), $\vec{u}_{in}$ the imposed velocity profile equal to the profile of a fully developed laminar flow through an annulus, $T_{in}$ the inlet temperature equal to 500 K, $T_{cath}$ the cathode temperature which is approximated by a Gaussian profile along the *z*-axis varying from 500 K at the inlet to 3000 K at the cathode tip, $h_w$ the convective heat transfer coefficient at the anode wall equal to $10^5$ W-m$^{-2}$-K$^{-1}$ (same value used in [21] and approximates the convective heat transfer due to turbulent flow in a pipe), and $T_w$ a reference cooling water temperature of 500 K. The boundary condition at the anode for $T_e$ has been used for its simplicity and to make the total heat transfer to the anode calculated with the NLTE and LTE models equivalent if $T_e \approx T_h$. $J_{cath}$ is the imposed current density over the cathode and is defined by:

$$J_{cath} = J_{cath0} \exp(-(r/R_c)^{n_c})$$ (32)

where $r$ is the radial distance from the torch axis ($r^2 = x^2 + y^2$), and $J_{cath0}$, $R_c$, and $n_c$ are parameters that specify the shape of the current density profile chosen to mimic the experimental measurements in [40]. The values of these parameters for the cases studied in Section 5 are: for 400 A, $J_{cath0} = 2 \cdot 10^8$ A-m$^{-2}$, $R_c = 0.81015$ mm, $n_c = 4$; and for 800 A, $J_{cath0} = 3 \cdot 10^8$ A-m$^{-2}$, $R_c = 0.91306$ mm, $n_c = 4$.

In the LTE model, due to the thermal equilibrium assumption, the electron temperature is equal to the heavy-particle temperature, which is low near the electrodes due to the strong cooling they experience. This causes the value of the equilibrium electrical conductivity near the electrodes to be extremely low (i.e. less than 0.01 S-m$^{-1}$), which limits the flow of electrical current to the electrodes. To allow the flow of electrical current from the plasma through the electrodes, the electrical conductivity is defined as $\sigma \leftarrow \max(\sigma_{eq}, 8 \cdot 10^3)$ ($\sigma, \sigma_{eq}$, and $8 \cdot 10^3$ in S-m$^{-1}$, and $\sigma_{eq}$ is the equilibrium electrical conductivity) in the region within 0.08 mm immediately adjacent to the electrodes. Similar approaches have been used extensively in the literature for the modeling of plasmas when the LTE assumption is invoked [10, 17-20, 22-23]. The NLTE model does not require any special treatment of the plasma-electrode boundary in order to allow the current flow, because the electron temperature will typically remain high in the arc attachment regions.

## 5. Modeling results

### 5.1. Simulation conditions

In this section we present results of our models applied to the simulation of the DC arc plasma torch shown in figure 6 working with pure argon, 60 slpm, 45° swirl injection, and either 400 or 800 A.

As explained in [10], simulations based on the LTE assumption and boundary conditions similar to those in Table 3 typically display excessively large voltage drops compared to experimental observations. More realistic voltage drops, and correspondingly arc lengths, are obtained when a *reattachment model* is included in the simulations. A reattachment model consists of a procedure that imitates the physical process in which a new arc attachment is formed (i.e. see [6-7]). Any reattachment model needs to answer the questions of, *where* and, *how* to produce the new attachment. The reattachment model developed in [10] answers these questions by: 1, a reattachment occurs at the location where the local electric field exceeds a certain pre-specified value $E_b$; and 2, at this location, a high electrical conductivity channel along the direction normal to the anode is introduced, inducing the formation of a new attachment. All the LTE simulations presented here have used the reattachment model reported in [10] with a value of $E_b = 5 \cdot 10^4$ V/m. As it will be seen in Section 5.4, this value of $E_b$ allows the matching of the obtained peak frequency of the LTE model with that of the NLTE model. Furthermore, the specification of an artificially high electrical conductivity in the regions immediately in front of the electrodes, as explained in Section 4, allows the formation of a new attachment if the arc gets "close enough" to the anode. More details of this type of arc reattachment are found in [10, 22-23].

### 5.2. Flow fields inside the torch

Figure 7 shows the instantaneous temperature distribution inside the torch operating with 400 A. The three-dimensionality of the flow can be observed as well as the effect of the cathode jet on the temperature distributions, an effect that is more noticeable in the NLTE results. The size and temperature of the anode spot, given by the distributions of $T$ and $T_h$ for the LTE and NLTE models respectively, are quite comparable. But, the size of the anode spot given by the $T_e$ distribution over the anode is significantly larger. This larger spot in part explains the lower resistance to the current flow in a NLTE model compared to a LTE model, as suggested in [10] and corroborated by the total voltage drop signals shown in Section 5.4. The location of the arc attachment for the LTE case cannot be observed clearly because the attachment is not crossed exactly by the vertical plane.

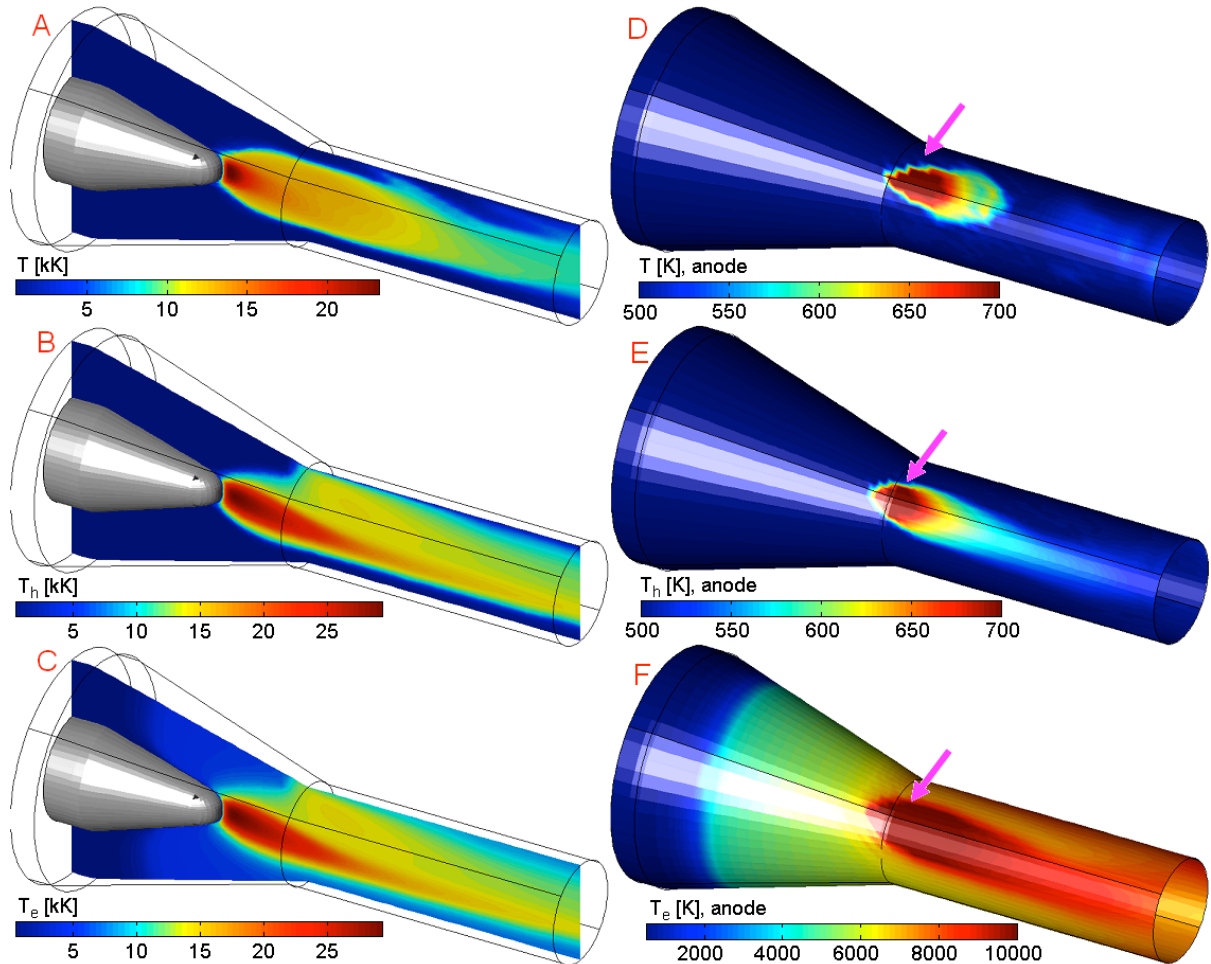

**Figure 7.** Instantaneous temperature distribution through a vertical plane across the torch (A, B, C) and close to the anode surface (D, E, F) for a torch operating with 400 A: Equilibrium temperature $T$, and non-equilibrium heavy-particle and electron temperatures, $T_h$ and $T_e$. The arrows indicate the location of the anode spot. The plots of $T_h$ and $T_e$ at the anode have been rotated 45° in the direction of the $z$-axis.

The marked difference between the temperatures obtained with the LTE and NLTE models suggest a significant difference in the energy distribution between the models. This difference is somewhat surprising considering that the NLTE model, as explained in Section 2.2, reduces to the LTE model if equilibrium conditions are assumed (i.e. if $T_h \approx T_e$ in the whole domain). It is reasonable to expect that this difference is due to at least two effects: (1) the boundary conditions used for the electron energy equation; and (2) the occurrence of a significant thermal non-equilibrium region within the domain. The first effect is clearly seen in the $T_e$ distribution over the anode in figure 7, which shows electron temperatures of $\sim 7000$ K in most of the boundary downstream of the anode attachment, and temperatures of $\sim 10000$ K at the location of the attachment. The second effect is observed in the distribution of the non-equilibrium parameter $\theta$, as presented in figure 8.

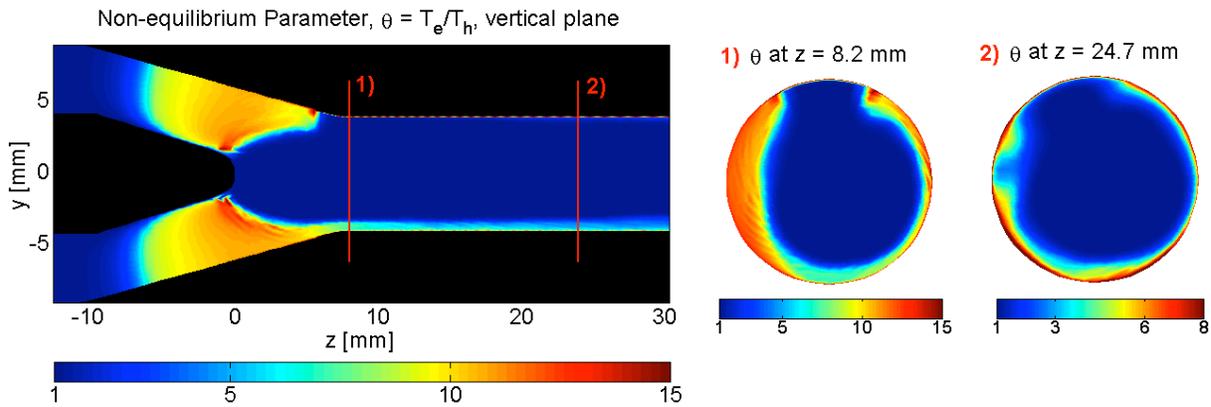

**Figure 8.** Non-equilibrium parameter $\theta$ for the same conditions of the plots in figure 7, NLTE.

Figure 8 shows that thermal non-equilibrium is dominant in the regions where the cold flow interacts with the plasma. The higher $\theta$ regions near the cathode tip and immediately upstream of the anode column seem to indicate that thermal non-equilibrium increases as the strength of the plasma – cold-flow interaction increases. The non-equilibrium region along the anode can also be observed at the opposite side of the anode attachment, in the region composing the cold flow boundary layer. The plots of $\theta$ at different cross sections indicate that thermal non-equilibrium in the cold boundary layer decreases as the flow progresses downstream, away from the arc (observe the different color scales used for cross section plots).

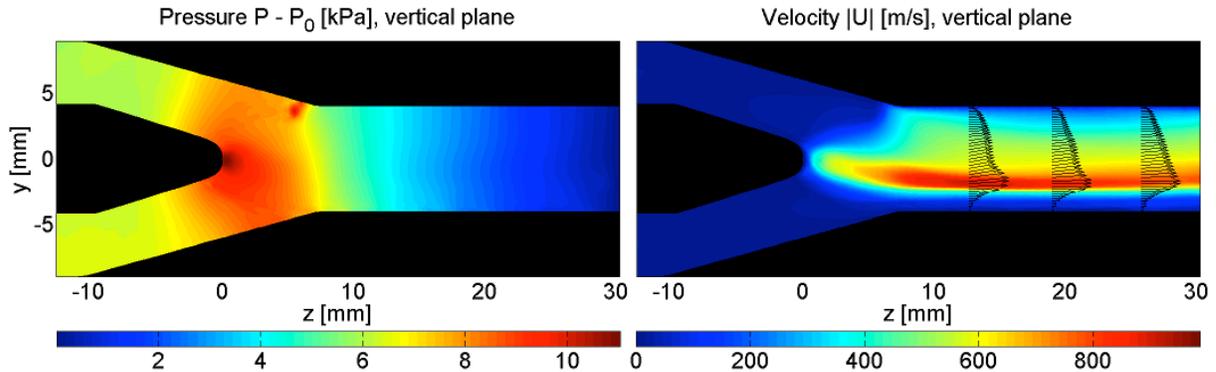

**Figure 9.** Pressure and velocity distribution for the same conditions of the plots in figure 7, NLTE.

The (over-)pressure and velocity distributions inside the torch for the NLTE model are shown in figure 9; the LTE results present similar characteristics. The pressure and velocity distributions clearly indicate the occurrence of a cathode jet: a high pressure region in front of the cathode tip due to the local constriction of the arc, producing electromagnetic pumping, and the corresponding formation of a high velocity region from the cathode tip outwards. The pressure distribution also shows a high pressure region immediately upstream of the anode column, which indicates that the cold flow tries to avoid entering the low-density arc. But, the flow that enters the plasma is rapidly accelerated, as seen in the velocity profiles in figure 9. Comparing the temperature distributions in figure 7 with the velocity distribution in figure 9, it seems evident that the temperature and velocity distributions are correlated, especially near the torch outlet. This fact has been observed experimentally in measurements of the plasma jet [8-9], and in previous simulations of plasma torches [22].

The velocity vectors in the velocity distribution plot indicate the velocity profiles in the $y$-$z$ plane ($\vec{u}(x=0, y, z)$) at different axial locations. These profiles show inflection points in the boundary between the plasma and the cold flow (i.e. see profile at $z \sim 12$ mm and $y \sim -3$ mm). As is well know from hydrodynamic stability theory [41], inflection points can trigger instabilities; particularly, the conditions of the flow around the arc column, as shown in figure 9, could trigger the Kelvin-Helmholtz instability [41]. This instability has not been observed in our simulations, probably because the spatial resolution used is not sufficient to capture the instability, or because the plasma flow naturally dampens these instabilities before they fully develop. Theoretical and experimental studies by Ragaller and collaborators [42-43] have revealed that the plasma – cold-flow interface is a strong vorticity-production region, which can favor the formation of fluid dynamic instabilities. Certainly the possible occurrence of fluid dynamic instabilities in the flow inside DC arc plasma torches requires further investigation.

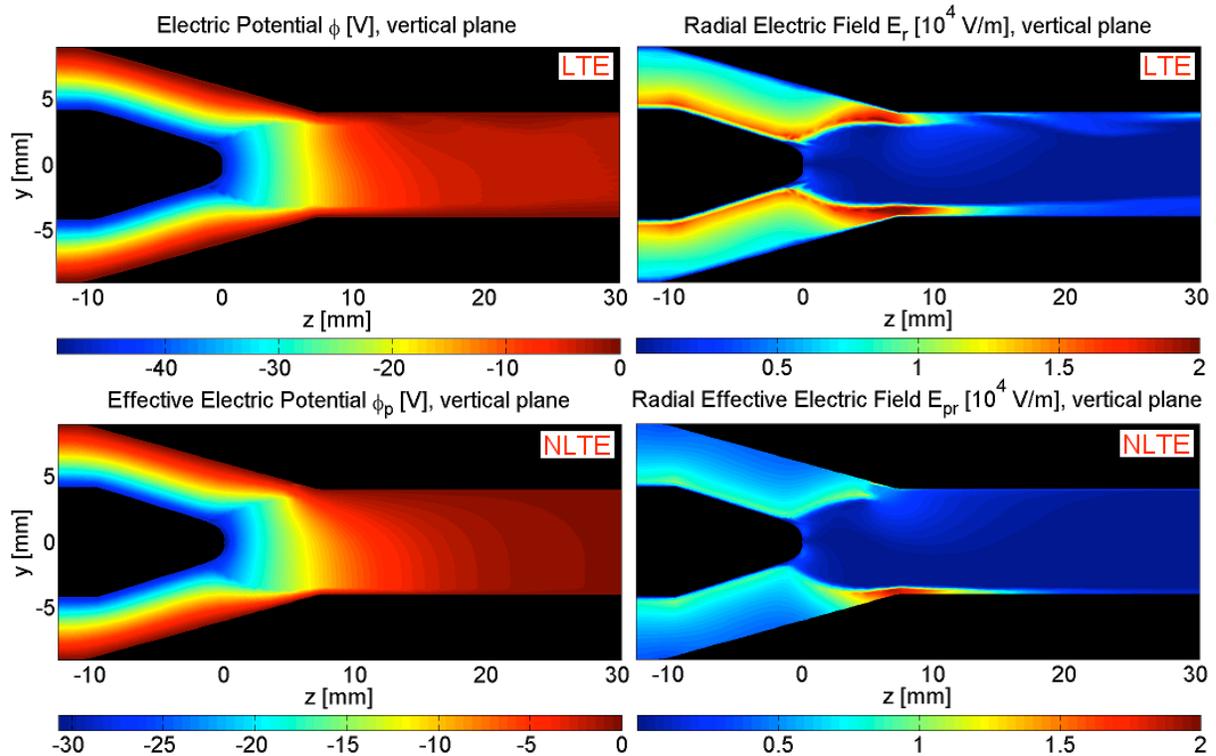

**Figure 10.** Electric potentials and fields for the LTE and NLTE models for the same conditions of the plots in figure 7.

Figure 10 presents the distributions of electric potential and field for the LTE model and of the effective electric potential and field for the NLTE model for the torch operating with 400 A. It is important to mention that the total voltage drop measured experimentally approximately corresponds, in the NLTE model, to the total voltage drop produced by the $\phi_p$ distribution (not by the $\phi$ distribution; see [15]), whereas in the LTE model both distributions are assumed equal. In figure 10 can be observed that, although the $\phi$ and $\phi_p$ distributions look alike, there is a marked difference between their values. This is remarkable considering that the locations of the arc attachment for both cases are comparable (see figure 7). This difference is more pronounced in the results presented in the next section.

Furthermore, it can be seen that the $E_p$ distribution shows lower values than the $E$ distribution (the same color scale has been used to emphasize their differences), especially around the cathode. The lower values of $E_p$ with respect to $E$ are due to the lower values of $\phi_p$ compared to $\phi$. But, the high values of $E_p$ at approximately the opposite side of the anode attachment (see figure 7 for the location of the attachment) will trigger the formation of a new attachment – without the use of a reattachment model – as it will be observed and described in the next section.

## 5.2. Arc reattachment processes

Figures 11 and 12 present snapshots of the reattachment process for the different cases studied showing the cathode jet, the movement of the anode column, the formation of a new arc attachment, the small fluctuation of the arc column, and the undulating behavior of the flow as it leaves the torch. As expected [1], the arc is more robust and has higher temperatures for the 800 A cases compared to the 400 A cases. It can also be observed how the anode attachment moves upstream initially; then, due to the net angular momentum over the arc, as explained in [22], the curvature of the arc increases, driving the arc towards the opposite side of the original attachment. These dynamics of the arc are interrupted by the occurrence of an arc reattachment process.

The formation of a new attachment in the NLTE model, as observed by the appearance of a small high temperature appendage (see arrows in figures 11 and 12), seems to be driven by high values of the local effective electric field (i.e. as seen in figure 10) and high values of electron temperature. Even though swirl injection is used, due to the relatively short time scale of the reattachment process, the arc reattaches at almost the opposite side of the original attachment. The reattachment process occurs in a natural manner mimicking the takeover and/or steady mode of operation of the torch (i.e. the results resemble the takeover mode for the 400 A case and an intermediate takeover-steady mode for the 800 A case). For the conditions studied, a reattachment model was not needed in the NLTE model. It is expected that non-equilibrium simulations of the restrike mode will require the use of a reattachment model similar to the one used in the LTE simulations in [10] or the one used in [25].

As explained in Section 5.1, in the LTE model a reattachment can occur either due to the application of the reattachment model (i.e. when the breakdown condition is satisfied) or due to the arc dynamics causing the arc to get "close enough" to the anode. The reattachment processes in the simulations of the LTE model shown in figure 11 and 12 are produced by the application of the reattachment model. The growth of a high temperature region from the arc column towards the anode can be observed in the LTE results in figures 11 and 12, which eventually initiates the formation of a new attachment. Moreover, the application of the reattachment model clearly disrupts the flow significantly.

Figure 13 shows the evolution of the total (effective) voltage drop across the torch during the reattachment processes in figures 11 and 12. The application of the reattachment model in the LTE simulations causes a relatively sharp decrease of the voltage drop (i.e. from point 2 to point 3 in the LTE results). In the NLTE simulations the reattachment process causes a more gradual decrease of the total voltage drop. The decrease in voltage drop between points 1 and 2 for the NLTE, 400 A, case is due to the movement of the anode attachment in the azimuthal direction and subsequent contraction of the arc, as observed in frames 1 and 2 of the NLTE results in figure 11.

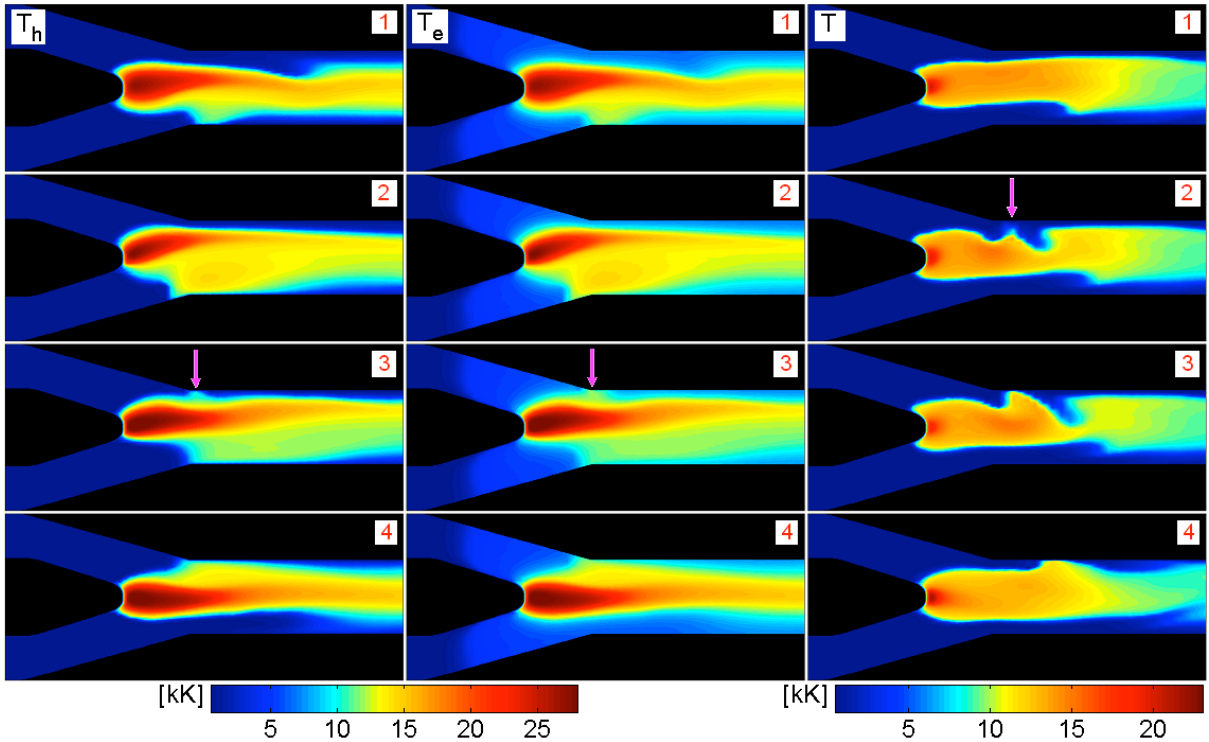

**Figure 11.** Temperature distributions inside the torch during the reattachment process for the torch operating with 400 A; NLTE ($T_h$ and $T_e$) and LTE ($T$) results. The arrows indicate the location where the new attachment starts forming. See figure 13 for the time instant of each frame.

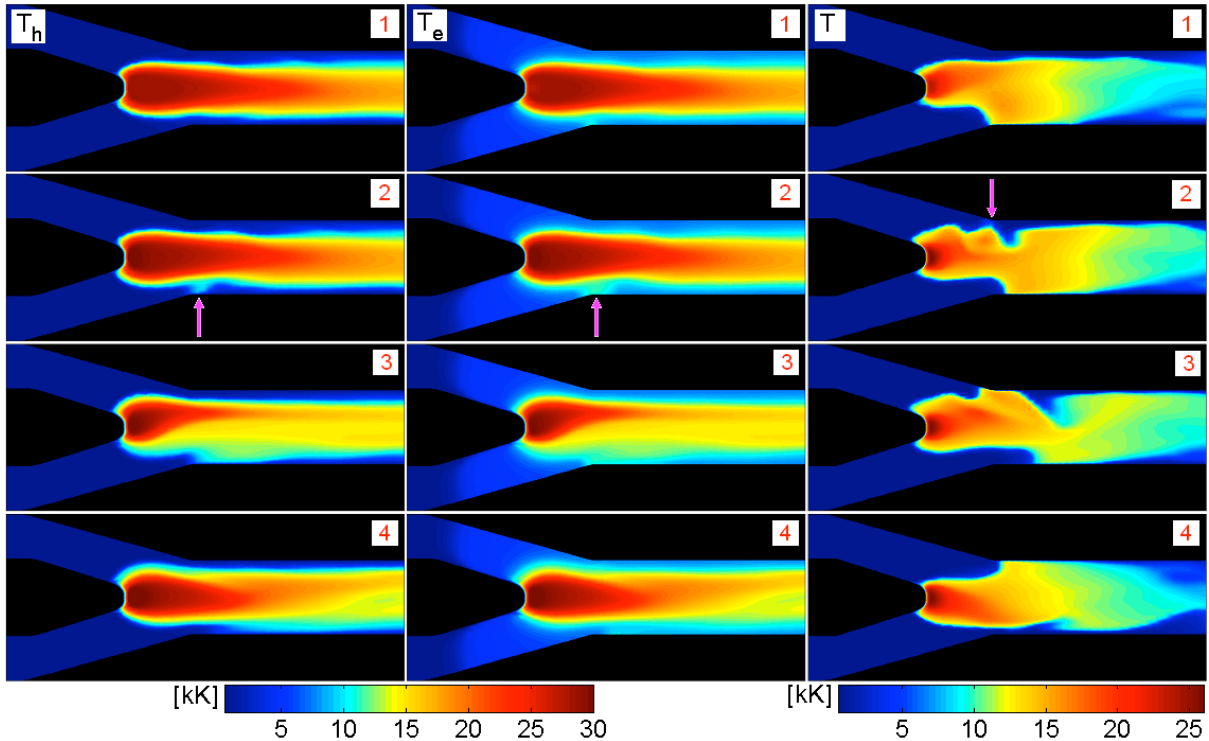

**Figure 12.** Temperature distributions inside the torch during the reattachment process for the torch operating with 800 A; NLTE ($T_h$ and $T_e$) and LTE ($T$) results. The arrows indicate the location where the new attachment starts forming. See figure 13 for the time instant of each frame.

Figure 13 also shows the clear difference between the values of voltage drop between the LTE and NLTE simulations. These are not only differences in the average voltage, but also in the amplitude of the voltage fluctuations; i.e. for the 800 A simulations, the reattachment process in the LTE model produces a decrease in the total voltage drop of ∼ 15 V, whereas in the NLTE model of only ∼ 7 V. The evolution of the voltage characteristics is analyzed in more detail in the next section.

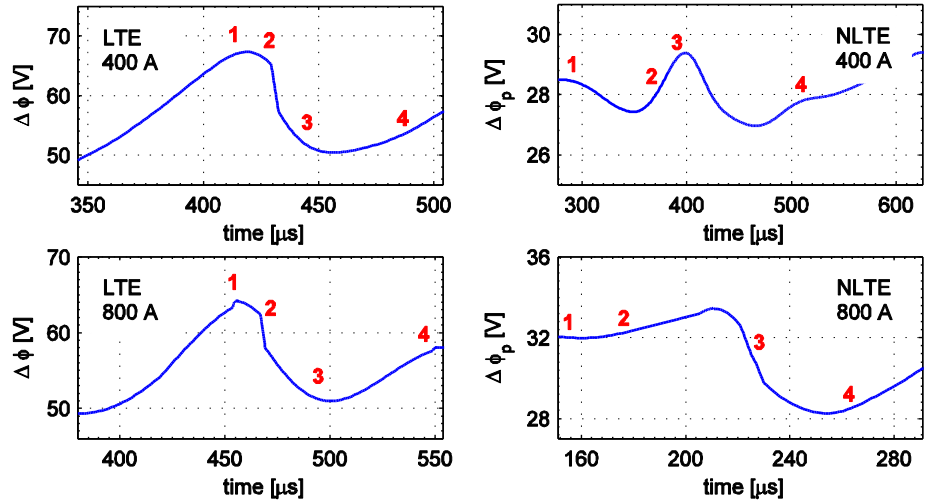

**Figure 13.** Voltage drop evolution during the reattachment process. The numbers indicate the time instant of the frames in figures 11 and 12.

### 5.3. Torch operational characteristics

Probably the most important operational characteristics of a plasma torch are the voltage drop signal and the outlet temperature and velocity profiles. These quantities are typically measured experimentally to diagnose the operation of the torch. Figure 14 shows the total voltage drop signals and their respective power spectra for the torch operating with 800 A, and the experimentally measured data in a SG-100 torch operating with Ar, flow rate of 60 slpm, 45° swirl injection, total current of 700 A, and using a new anode. Even though these operating conditions are typical in plasma spraying, instead of pure Ar, gas mixtures like Ar-He or Ar-H$_2$ are usually used due to their more suitable properties [1-3, 11]. Pure argon has been used in our model due to its relatively simple chemistry, but our current efforts are centered on applying our non-equilibrium model to gas mixtures.

As mentioned in [10], the use of a reattachment model allows the matching of the numerical results with experimental measurements of either, the peak frequency ($f_p$) or the magnitude of the voltage drop, but not both. This is also observed in the results in figure 14. The value of $E_b = 5 \cdot 10^4$ V/m makes the peak frequency of the LTE model match that of the NLTE model (both for 800 A). But, there are noticeable differences in the magnitudes of the total voltage drops. The larger voltage drops in the LTE model are likely due to the larger resistance to the current flow with respect to that in a NLTE model and what is observed in experiments. The lower resistance to the current flow in the NLTE model is due to the larger cross section of the arc (which is approximately identified by the regions with high electron temperature) and due to the lower electrical resistivity (= $1/\sigma$) caused by the high electron temperatures (considering that $\sigma$ is mostly a function of $T_e$). The larger cross section of the current path can be observed in the anode spot given by the electron temperature distribution in figure 7, whereas the higher electron temperatures are evident by comparing the $T_e$ and $T$ distributions in figures 11 and 12. Furthermore, the lack of adequate modeling of the plasma-electrode interfaces needs to be considered, which makes the obtained voltage drops from the LTE and NLTE models differ even more from those measured. But, the possible content of metal vapor near the electrodes could also explain the lower values of the voltage drops measured with respect to those obtained from the simulations.

The frequency plots show that the NLTE and LTE simulations have almost the same peak frequencies, whereas the experimental results show a broad frequency spectrum. This broad spectrum is typical of torches operating with a new anode [8], especially if pure argon and high currents are used. Although the broad spectrum of the experimental data does not permit to determine a clear peak frequency, the form of the experimental total voltage drop signal looks like the one obtained with the NLTE model, both resembling the takeover mode of operation [6-7]. Our NLTE simulations, as well as the numerical results in [19], indicate that the total voltage drop increases as the total current increases. Therefore, the larger values of voltage drop for the NLTE model compared to the experimental ones are probably due to the higher current used in the simulations.

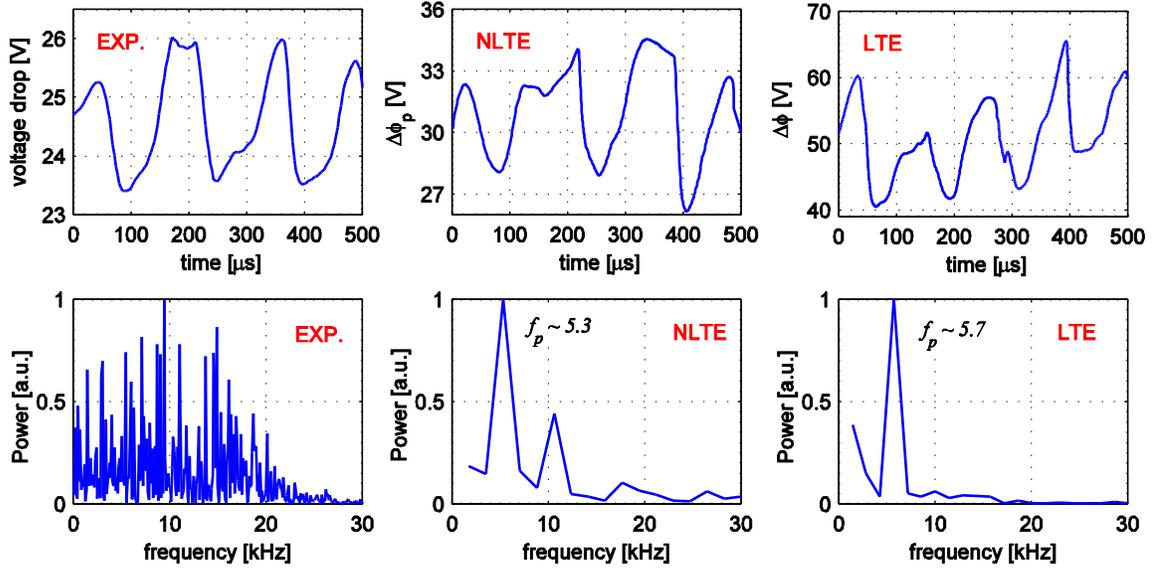

**Figure 14.** Experimental and numerical total voltage drop signals and their respective frequency spectra. The experimental conditions are: Ar, 60 slpm, and 700 A. The numerical results are for 800 A. The frequency spectrum of the experimental data has used data for a time interval of ~ 100 ms.

Figure 15 shows the average temperatures and velocities profiles at the torch outlet for the different cases studied. The outlet profiles are calculated as the averages between the time-averaged profiles along the $x$- and $y$-axis; i.e. the equilibrium temperature at the outlet $T_{out}$ in figure 15 is calculated by:

$$T_{out}(r) = \tfrac{1}{2}(\bar{T}(x=0, y, z=L) + \bar{T}(x, y=0, z=L)) \qquad (33)$$

where $L = 30.3$ mm is the axial length of the nozzle measured from the cathode tip (see figure 6), and the over-bar represents a time-averaged quantity; i.e. $\bar{T}$ is calculated by: $\bar{T} = \Delta_t^{-1} \int_0^{\Delta_t} T dt$ where $\Delta_t$ represents the total time interval. The total time interval used for the time averages is typically in the order of ~ 500 μs, which is very small compared with the intervals considered in experimental measurements (i.e. the experimental data in figure 14 have been obtained during a sampling period of

~ 100 ms, but with high temporal resolution). The profiles in figure 15 are not perfectly axi-symmetric (especially NLTE, 800 A) due to the limited extent of the time intervals used for their calculation, but are representative of the torch operation.

From figure 15, it is clear that the NLTE model produces higher temperatures and velocities at the outlet. This is not surprising due to the higher temperatures within the domain for the NLTE model, as discussed in Section 5.2. Moreover, the NLTE results show that the electron temperature is ~ 8000 K at the wall. This fact could suggest that there is a significant amount of current reaching the anode through the regions outside the anode spot. But, it needs to be considered that, due to the high degree of non-equilibrium close to the wall and the behavior of $\sigma$ in figure 3 at $T_e$ ~ 8000 K (see also [12]), that the electrical conductivity close to the wall is very small. Therefore very little current could flow through the regions outside the anode spot. A comparison with experimental results obtained for similar conditions [44] shows that the temperature distributions obtained with the NLTE model are close to the experimental ones (~ 12500 K for a torch operating with 400 A), whereas those obtained with the LTE model show lower temperatures.

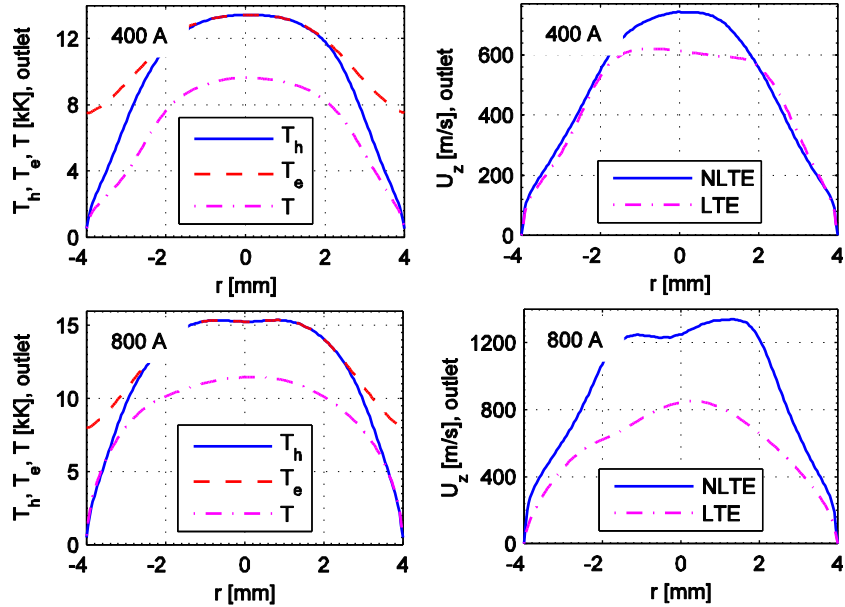

**Figure 15.** Average outlet temperatures and velocity profiles for the simulated cases.

**Table 4.** Heat to the plasma and thermal efficiency [a].

| *Model* | $I$ [A] | $\overline{Q}_{plasma}$ [kW] | $\Delta\overline{\phi}$ [V] | $\eta_{thermal}$ [%] |
|---------|---------|------------------------------|------------------------------|----------------------|
| LTE     | 400     | 3.23                         | 54.48                        | 14.8                 |
| NLTE    | 400     | 7.66                         | 27.23                        | 70.3                 |
| LTE     | 800     | 4.33                         | 49.76                        | 10.9                 |
| NLTE    | 800     | 18.30                        | 30.79                        | 74.2                 |

[a] flow rate = 60 slpm in al cases

The time-averaged total amount of heat transferred to the plasma $\overline{Q}_{plasma}$ and the thermal efficiency of the torch $\eta_{thermal}$ for the cases studied are displayed in Table 4. The heat to the plasma is approximated by $\overline{Q}_{plasma} \approx \dot{m}h_{ave}$, where $\dot{m}$ is the total mass flow rate and $h_{ave}$ is the average enthalpy at the outlet calculated from the temperature profiles in figure 15 (i.e. for the LTE model $h_{ave} = h(p_0, \overline{T})$). Although this definition of average enthalpy is not rigorous (i.e. $h_{ave}$ should be equal to $\bar{h}$ ), it reflects the enthalpy that would be calculated from experimentally measured time-averaged temperatures. The thermal efficiency is simply defined as $\eta_{thermal} = \bar{Q}_{plasma} /(I \cdot \Delta\bar{\phi})$ , where $I$ is the total current and $\Delta\bar{\phi}$ the time-averaged total (effective) voltage drop. This efficiency represents the fraction of electric energy that is used to heat the plasma. Considering that thermal efficiencies are typically in the range between 40 and 60% (i.e. see [1, 19]), the results in Table 9 indicate that the thermal efficiency is underestimated in the LTE simulations and overestimated in NLTE ones. The low efficiencies of the LTE simulations are mainly due to the large total voltage drops obtained (~ 2 times larger than those experimentally found) and also due to the underestimation of the heat transferred to the plasma. The high efficiencies of the NLTE simulations indicate that the amount of heat transferred to the electrodes, particularly to the anode, is underestimated. The underestimation of the heat transferred to the anode is likely due to the simplified boundary conditions used for the energy equations (see Table 3), which neglect the effects of electron condensation, electron enthalpy flow, ion recombination, and anode fall. Particularly, the thermal efficiencies obtained with the NLTE models could be reduced if electron condensation, which can account for up to 60% of the total amount of heat transferred to the anode [15], is included.

## 6. Conclusions

A two-temperature thermal non-equilibrium model has been developed and applied to the three-dimensional and time-dependent simulation of the flow inside a DC arc plasma torch operating with argon for different total currents. The non-equilibrium model has been compared with a Local Thermodynamic Equilibrium model, which is equivalent to the non-equilibrium model if thermal equilibrium is assumed. The fluid and electromagnetic equations of both models have been approximated numerically in a fully-coupled approach by a variational multi-scale finite element method. The results show significant differences between the equilibrium and non-equilibrium models; the results of the latter show better agreement with experimental observations. Particularly, the non-equilibrium results display higher temperatures and velocities, especially at the torch outlet, and total voltage drops lower than those of the equilibrium model. The lower total voltage drops obtained with the non-equilibrium model are due to its lower resistance to the current flow because of its higher electron temperatures and larger size of the anode spot, as given by the electron temperature distribution over the anode. Although the total voltage drops of the non-equilibrium results are still slightly higher than those measured, the overall shape of the voltage signal resembles that measured experimentally, indicating the takeover mode of operation of the torch. Furthermore, the non-equilibrium model did not need a separate model to produce a reattachment process which limits the magnitude of the total voltage drop and arc length, whereas the equilibrium model did. But, it is expected that non-equilibrium simulations of the restrike mode of operation will need a breakdown model to produce more realistic results. Moreover, the non-equilibrium results show large non-equilibrium regions near the cold-flow – plasma interaction region and in front of the anode surface. The simulations also revealed that the level of non-equilibrium increases as the plasma – cold-flow interaction increases. This effect, added to the boundary conditions used in the non-equilibrium model, has a marked effect on the overall energy balance of the flow.

## Acknowledgments

This research has been partially supported by NSF Grant CTS-0225962. Computing time from a grant from the University of Minnesota Supercomputing Institute and support for the first author by a Doctoral Dissertation Fellowship from the University of Minnesota are gratefully acknowledged. The authors would like to thank David Outcalt for providing the experimental data, and the anonymous referees whose valuable comments have improved the quality of this paper.